\documentclass[aps, prl, twocolumn, reprent, superscriptaddress]{revtex4-1}
\newcommand{\EQ}[1]{Eq.~(\ref{#1})}
\newcommand{\FIG}{Fig.~}
\newcommand{\GZ}[2][]{g^{#1}(#2)}
\newcommand{\HZ}[2][]{\Theta^{#1}(#2)}
\newcommand{\AM}[1][]{\langle a^{#1}\rangle}
\newcommand{\SM}{\sqrt{m}}
\newcommand{\SINH}[1][]{\sinh^{#1}(\beta\SM\tau)}
\newcommand{\COSH}[1][]{\cosh^{#1}(\beta\SM\tau)}
\newcommand{\COSHM}{(\COSH-1)}
\newcommand{\mat}[1]{\bm{#1}}
 
\newcommand{\add}[1]{#1}
\newcommand{\del}[1]{}
\usepackage{color}
\usepackage[normalem]{ulem}

\usepackage{amsmath}
\usepackage{bm}
\usepackage{graphicx}
\usepackage{epstopdf}

\begin{document}
\title{Concurrency-induced transitions in epidemic dynamics on temporal networks}
\author{Tomokatsu Onaga}
\affiliation{Department of Physics, Kyoto University, Kyoto 606-8502, Japan}
\affiliation{MACSI, Department of Mathematics and Statistics, University of Limerick, Limerick V94 T9PX, Ireland}
\author{James P. Gleeson}
\affiliation{MACSI, Department of Mathematics and Statistics, University of Limerick, Limerick V94 T9PX, Ireland}
\author{Naoki Masuda}
\email{naoki.masuda@bristol.ac.uk}
\affiliation{Department of Engineering Mathematics, University of Bristol, Woodland Road, Bristol BS8 1UB, United Kingdom}
\date{\today}
\begin{abstract}
Social contact networks underlying epidemic processes in humans and animals are highly dynamic.
 The spreading of infections on such temporal networks can differ dramatically from spreading on static networks. 
 We theoretically investigate the effects of concurrency, the number of neighbors that a node has at a given time point, on the epidemic threshold in the stochastic susceptible-infected-susceptible dynamics on temporal network models. We show that network dynamics can suppress epidemics (i.e., yield a higher epidemic threshold) when the nodes' concurrency is low, but can also enhance epidemics when the concurrency is high. We analytically determine different phases of this concurrency-induced transition, and confirm our results with numerical simulations.
\end{abstract}
\maketitle

\noindent\emph{Introduction:}
Social contact networks---on which infectious diseases occur in humans and animals or viral information spreads online and offline---are mostly dynamic. Switching of partners and the (usually non-Markovian) activity of individuals, for example, shape network dynamics on such temporal networks~\cite{Holme2012,Holme2015EurPhysJB,MasudaLambiotte2016book}. A better understanding of epidemic dynamics on temporal networks is needed to help  improve predictions of, and interventions in, emergent infectious diseases, to design vaccination strategies, and to identify viral marketing opportunities.  This is particularly so because what we know about epidemic processes on static networks~\cite{Keeling2005JRSocInterface,Barrat2008book,Pastor-Satorras2015,Porter2016} is only valid when the time scales of the network dynamics and of the infectious processes are well separated. In fact, temporal properties of networks, such as long-tailed distributions of intercontact times, temporal and cross-edge correlation in intercontact times, and entries and exits of nodes, considerably alter how infections spread in a network~\cite{Bansal2010JBiolDyn,Holme2012,Masuda2013F1000,Holme2015EurPhysJB,MasudaLambiotte2016book}.

In the present study, we focus on a relatively neglected component of temporal networks, i.e., the number of concurrent contacts that a node has. Even if two temporal networks are the same when aggregated over a time horizon, they may be different as temporal networks due to different levels of concurrency. Concurrency is a longstanding concept in epidemiology, in particular in the context of monogamy or polygamy affecting sexually transmitted infections~\cite{Morris1995,Kretzschmar1996a,Miller2016}. Modeling studies to date largely agree that a level of high concurrency (e.g., polygamy as opposed to monogamy) enhances epidemic spreading in a population. However, this finding, while intuitive, lacks theoretical underpinning. First, some models assume that the mean degree, or equivalently the average contact rate, of nodes increases as the concurrency increases~\cite{Watts1992,Eames2004,Doherty2006,Gurski2016}. In these cases, the observed enhancement in epidemic spreading is an obvious outcome of a higher density of edges rather than a high concurrency. Second, other models that vary the level of concurrency while preserving the mean degree are numerical~\cite{Morris1995,Kretzschmar1996a,Bauch2000,Leung2015}. In the present study, we \del{build on} \add{use} the analytically tractable activity-driven model of temporal networks~\cite{Perra2012,Starnini2013PhysRevE,Ribeiro2013a,Liu2014,Zino2016} to explicitly modulate the size of the concurrently active network with the structure of the aggregate network fixed. With this machinery, \add{we carefully treat extinction effects, derive an analytically tractable matrix equation using a probability generating function for dynamical networks, and reveal non-monotonic effects of link concurrency on spreading dynamics.} \del{we} \add{We} show that the dynamics of networks can either enhance or suppress infection, depending on the amount of concurrency that individual nodes have. Note that analysis of epidemic processes driven by discrete pairwise contact events, which is a popular approach ~\cite{Holme2012,Holme2015EurPhysJB,MasudaLambiotte2016book,Masuda2013F1000,Zino2016,VazquezA2007PhysRevLett,Karsai2011PhysRevE, Miritello2011PhysRevE,Stehle2011BMCMed}, does not address the problem of concurrency because we must be able to control the number of simultaneously active links possessed by a node in order to examine the role of concurrency without confounding with other aspects.

\begin{figure}
\centering
\includegraphics[width=3.4in]{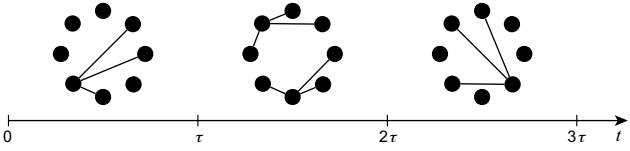}
\caption{Schematic of an activity-driven network with $m=3$.}
\label{f1}
\end{figure}

\noindent\emph{Model:}
We consider the following continuous-time susceptible-infected-susceptible (SIS) model on a discrete-time variant of activity-driven networks, which is a generative model of temporal networks~\cite{Perra2012,Starnini2013PhysRevE,Ribeiro2013a,Liu2014,Zino2016}. The number of nodes is denoted by $N$. Each node $i$~$(1\le i\le N)$ is assigned an activity potential $a_i$, drawn from a probability density $F(a)$~$(0<a\le1)$. Activity potential $a_i$ is the probability with which node $i$ is activated in a window of constant duration $\tau$. If activated, node $i$ creates $m$ undirected links each of which connects to a randomly selected node (\FIG\ref{f1}). If two nodes are activated and send edges to each other, we only create one edge between them. However, for large $N$ and relatively small $a_i$, such events rarely occur. After a fixed time $\tau$, all edges are discarded. Then,  in the next time window, each node is again activated with probability $a_i$, independently of the activity in the previous time window, and connects to randomly selected nodes by $m$ undirected links. We repeat this procedure. Therefore, the network changes from one time window to another and is an example of a switching network~\cite{Liberzon2003,Masuda2013,Hasler2013a,Speidel2016}. A large $\tau$ implies that network dynamics are slow compared to epidemic dynamics. In the limit of $\tau\to 0$, the network blinks infinitesimally fast, enabling the dynamical process to be approximated on a time-averaged static network, as in~\cite{Hasler2013a}.

For the SIS dynamics, each node takes either the susceptible or infected state. At any time, each susceptible node contracts infection at rate $\beta$ per infected neighboring node. Each infected node recovers at rate $\mu$ irrespectively of the neighbors' states. Changing $\tau$ to $c\tau$ $(c>0)$ is equivalent to changing $\beta$ and $\mu$ to $\beta/c$ and $\mu/c$, respectively, while leaving $\tau$ unchanged. Therefore, we set $\mu=1$ without loss of generality.

\noindent\emph{Analysis:}
We calculate the epidemic threshold as follows. \add{First, we formulate SIS dynamics near the epidemic threshold on a static star graph, which is the building block of the activity-driven model, while explicitly considering extinction effects. Second, we convert the obtained set of linear difference equations into a tractable mathematical form with the use of a probability generating function of an activity distribution. Third, the epidemic threshold is obtained from an implicit function.} For the sake of the analysis, we assume that star graphs generated by an activated node, which we call the hub, are disjoint from each other. Because a star graph with hub node $i$ overlaps with another star graph with probability $\approx m\sum_{j\ne i} a_j(m+1)/N\propto m^2\AM$, where $\AM\equiv\int {\rm d}aF(a)a$ is the mean activity potential, we impose $m^2\AM\ll1$. \add{(However, our method works better than the so-called individual-based approximation even when $m^2 \langle a \rangle = 0.5$, as shown in the Supplemental Material.)}
We denote by $\rho(a,t)$ the probability that a node with activity $a$ is infected at time $t$. The fraction of infected nodes in the entire network at time $t$ is given by $\langle \rho(t) \rangle\equiv\int {\rm d}aF(a)\rho(a,t)$. Let $c_1$ be the probability with which the hub in an isolated star graph is infected at time $t+\tau$, when the hub is the only infected node at time $t$ and the network has switched to a new configuration right at time $t$. Let $c_2$ be the probability with which the hub is infected at $t+\tau$ when only a single leaf node is infected at $t$. The probability that a hub with activity potential $a$ is infected after the duration $\tau$ of the star graph, denoted by $\rho_1$, is given by
\begin{equation}
  \label{eq:rho1}
  \rho_1(a,t+\tau) = c_1\rho(a,t) + c_2m\langle\rho(t) \rangle.
\end{equation}
In deriving \EQ{eq:rho1}, we considered the situation near the epidemic threshold such that at most one node is infected in the star graph at time $t$ [and hence $\rho(a,t), \langle \rho(t)\rangle \ll 1$].
The probability that a leaf with activity potential $a$ that has a hub neighbor with activity potential $a'$ is infected after time $\tau$ is analogously given by
\begin{equation}
  \label{eq:rho2}
  \rho_2(a,a',t+\tau) = c_3\rho(a,t) +  c_4\rho(a',t) + c_5(m-1)\langle\rho(t) \rangle,
\end{equation}
where $c_3$, $c_4$, and $c_5$ are the probabilities with which a leaf node with activity potential $a$ is infected after duration $\tau$ when only that leaf node, the hub, and a different leaf node is infected at time $t$, respectively. We derive formulas for $c_i$ $(1\le i\le 5)$ in the Supplemental Material. The probability that an isolated node with activity potential $a$ is infected after time $\tau$ is given by $e^{-\tau}\rho(a,t)$. By combining these contributions, we obtain
\begin{eqnarray}
  \label{eq:rho}
  \rho(a, t+\tau) &=& a\rho_1(a, t+\tau) + \int {\rm d}a'F(a')ma'\rho_2(a,a', t+\tau) \nonumber\\
  &+&(1-a-m\AM)e^{-\tau}\rho(a,t).
\end{eqnarray}

To analyze \EQ{eq:rho} further, we take a generating function approach. \add{With this approach, one trades a probability distribution for a probability generating function whose derivatives provide us with useful information about the distribution such as its moments. Furthermore, it often makes analysis easier, in particular linear analysis.} By multiplying \EQ{eq:rho} by $z^a$ and averaging over $a$, we obtain
\begin{eqnarray}
  \HZ{z, t+\tau} &=& c_1'\HZ[(1)]{z,t}
  +c_2'\HZ{1,t}\GZ[(1)]{z}+c_3'\HZ{z,t} \nonumber \\
                 &+& \left[c_4'\HZ[(1)]{1,t} +c_5'\HZ{1,t}\right]\GZ{z},
  \label{eq:Theta}
\end{eqnarray}
where $c_1^\prime\equiv c_1-e^{-\tau}$, $c_2^\prime\equiv mc_2$, $c_3^\prime\equiv e^{-\tau}+m\AM(c_3-e^{-\tau})$, $c_4^\prime\equiv mc_4$, $c_5^\prime\equiv m(m-1)\AM c_5$, $\GZ{z}\equiv\int {\rm d}aF(a)z^a$ is the probability generating function of $a$, $\HZ{z,t}\equiv\int {\rm d}aF(a)\rho(a,t)z^a$, and throughout the paper the superscript $(n)$ represents the $n$th derivative with respect to $\ln z$. \add{Although \EQ{eq:rho} is an infinite dimensional system of linear difference equations, \EQ{eq:Theta} is a single difference equation of $\HZ{z,t}$ and its derivative~\cite{Silk2014}.}

We expand $\rho(a,t)$ as a Maclaurin series as follows:
\begin{equation}
\rho(a,t)=\sum_{n=1}^\infty w_n(t)a^{n-1}.
\end{equation}
\add{Using this polynomial basis representation (the convergence is proven in the Supplemental Material), we can consider the differentiations in \EQ{eq:Theta} (i.e., $\HZ[(1)]{z,t}$ and $g^{(1)}(z)$) as an exchange of bases and convert \EQ{eq:Theta} into a tractable matrix form.} Let $p_0$ be the fraction of initially infected nodes, which are selected uniformly at random, independently of $a$. We represent the initial condition as $\bm w(t=0) \equiv (w_1(0), w_2(0), \ldots)^{\top} = \left(p_0, 0, 0, \ldots\right)^{\top}$. Epidemic dynamics near the epidemic threshold obey linear dynamics given by
\begin{equation}
  \bm w(t+\tau)=\mat T(\tau)\bm w(t).
\end{equation}
By substituting $\HZ{z,t} = \sum_{n=1}^\infty w_{n}(t)\GZ[(n-1)]{z}$ and $\GZ[(n-1)]{1}=\AM[n-1]$ in \EQ{eq:Theta}, we obtain
\begin{widetext}
\begin{equation}
  \mat T =
  \begin{pmatrix}
    c_3'+\AM c_4'+c_5' & \AM[2] c_4'+\AM c_5' & \AM[3] c_4'+\AM[2] c_5' & \AM[4] c_4'+\AM[3] c_5' & \AM[5] c_4'+\AM[4] c_5' & \cdots \\
    c_1'+c_2' & \AM c_2'+ c_3' & \AM[2] c_2' & \AM[3] c_2' & \AM[4] c_2' & \cdots \\
    0 & c_1' & c_3' & 0 & 0 & \cdots \\
    0 & 0 & c_1' & c_3' & 0 & \cdots \\
    0 & 0 & 0 & c_1' & c_3' & \cdots \\
    \vdots & \vdots & \vdots & \vdots & \vdots & \ddots
  \end{pmatrix}.
  \label{eq:T}
\end{equation}
\end{widetext}
 A positive prevalence $\left< \rho(t)\right>$ (i.e., a positive fraction of infected nodes in the equilibrium state) occurs only if the largest eigenvalue of $\mat T(\tau)$ exceeds $1$\add{, because in this situation the probability of being infected grows in time, at least in the linear regime}. Therefore, we get the following implicit function for the epidemic threshold, denoted by $\beta_{\rm c}$:
\begin{eqnarray}
  f(\tau, \beta_{\rm c})&\equiv&\frac{(1-r)(1-s)-(1+q)u}{S(q)} \nonumber\\
  &-&qr-qs+qrs-q^2u-rs=0, \label{eq:implicit eq}
\end{eqnarray}
where $S(q) \equiv \sum_{n=0}^\infty (\AM[n+2]/\AM^{n+2})q^{n}=(1/\AM^2)\left\{\left\langle (a^2)/[1-(a/\AM)q] \right\rangle\right\}$, $q\equiv\AM c_1'/(1-c_3')$, $r\equiv\AM c_2'/(1-c_3')$, $s\equiv\AM c_4'/(1-c_3')$, and $u\equiv c_5'/(1-c_3')$ (see Supplemental Material for the derivation). Note that $f$ is a function of $\beta$ ($=\beta_{\rm c}$) through $q$, $r$, $s$, and $u$, which are functions of $\beta$. In general, we obtain $\beta_{\rm c}$ by numerically solving \EQ{eq:implicit eq}, but some special cases can be determined analytically.

In the limit $\tau\to0$, \EQ{eq:implicit eq} gives $\beta_{\rm c}=\left[  m\left(\AM+\sqrt{\AM[2]}\right)\right]^{-1}$, which coincides with the epidemic threshold for the activity-driven model derived in the previous studies~\cite{Perra2012, Liu2014}. In fact, this $\beta_{\rm c}$ value is the epidemic threshold for the aggregate (and hence static) network, whose adjacency matrix is given by $A_{ij}^* \approx m(a_i+a_j)/N$ ~\cite{Speidel2016, MasudaLambiotte2016book}, as demonstrated in \FIG{S1}.

For general $\tau$, if all nodes have the same activity potential $a$, and if $m=1$,  we obtain $\beta_{\rm c}$ as the solution of the following implicit equation:
\begin{eqnarray}
  &&2ae^\frac{\left(\beta_{\rm c}-1\right)\tau}{2}\left[\cosh\left(\frac{\kappa_{\rm c}\tau}{2}\right) +\frac{1+3\beta_{\rm c}}{\kappa_{\rm c}}\sinh\left(\frac{-\kappa_{\rm c}\tau}{2}\right) \right] \nonumber \\
  &-&e^\tau+1-2a=0, \label{eq:et for 1}
  \end{eqnarray}
where $\kappa_{\rm c}=\sqrt{\beta_{\rm c}^2+6\beta_{\rm c}+1}$.

\begin{figure}
\centering
\includegraphics[width=3.4in]{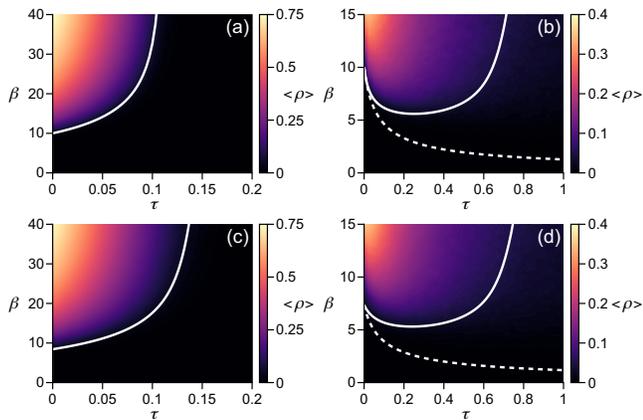}
\caption{Epidemic threshold and the numerically-simulated prevalence when $m=1$ (a),(c)\del{, and (e)} and $m=10$ (b),(d)\del{, and (f)}. In (a) and (b), all nodes have the same activity potential value $a$. The solid lines represent the analytical estimate of the epidemic threshold [\EQ{eq:implicit eq}; we plot \EQ{eq:et for 1} instead in (a)]. The dashed lines represent the epidemic threshold obtained from the individual-based approximation (Supplementa\add{l}\del{ry} Material). The color indicates the prevalence. In (c) and (d), the activity potential ($\epsilon\le a_i\le0.9$, $1\le i\le N$) obeys a power-law distribution with exponent $3$. In (a)--(d), we set $N=2000$ and adjust the values of $a$ and $\epsilon$ such that the mean degree is the same ($\langle  k \rangle=0.1$) in the four cases. \del{In (e) and (f), the activity potential is constructed from workplace contact data obtained from the SocioPatterns project~\cite{Genois2015}. This data set contains contacts between pairs of $N=92$ individuals measured every $20$ seconds. We calculate the degree of each node averaged over time, denoted by $\langle k_i \rangle$, and define the activity potential as $a_i=\left[\langle k_i \rangle-\langle k \rangle/2 \right]/m$.} We simulate the stochastic SIS dynamics using the quasistationary state method~\cite{DeOliveira2005}, as in~\cite{Speidel2016}, and calculate the prevalence averaged over $100$ realizations after discarding the first $15\,000$ time steps. We set the step size $\Delta t=0.002$\del{ in (a)--(d) and $\Delta t=0.001$ in (e) and (f)}. \add{Qualitatively similar results are obtained for the variant of the activity-driven model with a reinforcement mechanism of link creation~\cite{Karsai2014} (\FIG{S3}).}}
\label{f2}
\end{figure}

The theoretical estimate of the epidemic threshold [\EQ{eq:implicit eq}; we use \EQ{eq:et for 1} in the case of $m=1$] is shown by the solid lines in Figs. \ref{f2}(a) and \ref{f2}(b). It is compared with numerically calculated prevalence values for various $\tau$ and $\beta$ values shown in different colors. Equations (\ref{eq:implicit eq}) and (\ref{eq:et for 1}) describe the numerical results fairly well. When $m=1$, the epidemic threshold increases with $\tau$ and diverges at $\tau\approx0.1$~[\FIG{\ref{f2}}(a)]. Furthermore, slower network dynamics (i.e.,  larger values of $\tau$) reduce the prevalence for all values of $\beta$. In contrast, when $m=10$, the epidemic threshold decreases and then increases as $\tau$ increases~[\FIG{\ref{f2}}(b)]. The network dynamics (i.e., finite $\tau$) impact epidemic dynamics in a qualitatively different manner depending on $m$, i.e., the number of concurrent neighbors that a hub has. Note that the estimate of $\beta_{\rm c}$ by the individual-based approximation~(\cite{Speidel2016}, see Supplemental Material for the derivation), which may be justified when $m\gg1$, is consistent with the numerical results and our theoretical results only at small $\tau$ [a dashed line in \FIG{\ref{f2}}(b)]. \del{The results shown in Figs. \ref{f2}(a) and \ref{f2}(b) are q}\add{Q}ualitatively similar results are found, when the activity potential $a$ is power-law distributed [Figs. \ref{f2}(c) and \ref{f2}(d)]\del{ and when $F(a)$ is constructed from empirical data obtained from the SocioPatterns project~\cite{Genois2015} (Figs. \ref{f2}(e) and \ref{f2}(f))}.

\begin{figure}
\centering
\includegraphics[width=3.4in]{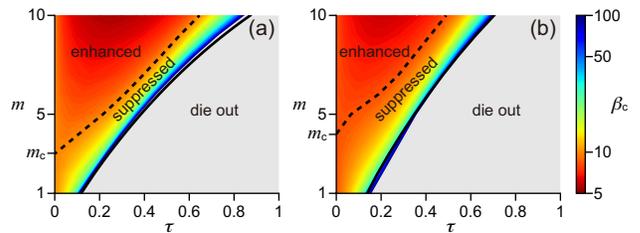}
\caption{Phase diagrams for the epidemic threshold, $\beta_{\rm c}$, when the activity potential is (a) equal to $a$ for all nodes, or (b) obeys a power-law distribution with exponent $3$~($\epsilon\le a_i\le0.9$). We set $\langle k \rangle=0.1$ at $m=1$ and adjust the value of $a$ and $\epsilon$ such that $\beta_{\rm c}$ takes the same value for all $m$ at $\tau=0$. In the ``die out'' phase, infection eventually dies out for any finite $\beta$. In the ``suppressed'' phase, $\beta_{\rm c}$ is larger than the $\beta_{\rm c}$ value at $\tau=0$. In the ``enhanced'' phase, $\beta_{\rm c}$ is smaller than the $\beta_{\rm c}$ value at $\tau=0$. The solid and dashed lines represent $\tau_*$~[\EQ{eq:tau star}] and $\tau_{\rm c}$, respectively. The color bar indicates the $\beta_{\rm c}$ values. In the gray regions, $\beta_{\rm c}>100$.}
\label{f3}
\end{figure}

To illuminate the qualitatively different behaviors of the epidemic threshold as $\tau$ increases, we determine a phase diagram for the epidemic threshold. We focus our analysis on the case in which all nodes share the activity potential value $a$, noting that qualitatively similar results are also found for power-law distributed activity potentials [\FIG{\ref{f3}}(b)]. We calculate the two boundaries partitioning different phases as follows. First, we observe that the epidemic threshold diverges \del{for $\tau > \tau_*$} \add{at $\tau = \tau_*$}. In the limit $\beta \to\infty$, infection starting from a single infected node in a star graph immediately spreads to the entire star graph, leading to $c_i\to1$ $(1\le i\le5)$. By substituting $c_i\to1$ in \EQ{eq:implicit eq}, we obtain $f(\tau_*, \beta_{\rm c} \to\infty) = 0$, where
\begin{equation} \tau_*=\ln\frac{1-(1+m)a}{1-(1+m)^2a}.
  \label{eq:tau star}
\end{equation}
When $\tau>\tau_*$, infection always dies out even if the infection rate is infinitely large. This is because, in a finite network, infection always dies out after a sufficiently long time due to stochasticity~\add{\cite{Keeling2008,Simon2011,Hindes2016}}. Second, although $\beta_{\rm c}$ eventually diverges as $\tau$ increases, there may exist $\tau_{\rm c}$ such that $\beta_{\rm c}$ at $\tau<\tau_{\rm c}$ is smaller than the $\beta_{\rm c}$ value at $\tau=0$. Motivated by the comparison between the behavior of $\beta_{\rm c}$ at $m=1$ and $m=10$~(\FIG{\ref{f2}}), we postulate that $\tau_{\rm c}$ ($>0$) exists only for $m>m_{\rm c}$. Then, we obtain ${\rm d}\beta_{\rm c}/{\rm d}\tau=0$ at $(\tau, m)=(0, m_{\rm c})$. The derivative of \EQ{eq:implicit eq} gives $\partial f/\partial \tau+(\partial f/\partial\beta_{\rm c})({\rm d}\beta_{\rm c}/{\rm d}\tau)=0$.
Because ${\rm d}\beta_{\rm c}/{\rm d}\tau=0$ at $(\tau,m)=(0,m_{\rm c})$, we obtain $\partial f / \partial \tau = 0$, which leads to
\begin{equation}
  m_{\rm c}=\frac{3}{1-4a}.
  \label{eq:mc}
\end{equation}
When $m < m_{\rm c}$, network dynamics (i.e., finite $\tau$) always reduce the prevalence for any $\tau$ [Figs. \ref{f2}(a) and \ref{f2}(c)]. When $m > m_{\rm c}$, a small $\tau$ raises the prevalence as compared to $\tau=0$ (i.e., static network) but a larger $\tau$ reduces the prevalence [Figs. \ref{f2}(b) and \ref{f2}(d)].

The phase diagram based on Eqs. (\ref{eq:tau star}) and (\ref{eq:mc}) is shown in \FIG{\ref{f3}}(a). The $\beta_{\rm c}$ values numerically calculated by solving \EQ{eq:implicit eq} are also shown in the figure. It should be noted that the parameter values are normalized such that $\beta_{\rm c}$ has the same value for all $m$ at $\tau=0$. We find that the dynamics of the network may either increase or decrease the prevalence, depending on the number of connections that a node can simultaneously have, extending the results shown in \FIG{\ref{f2}}.

These results are not specific to the activity-driven model. The phase diagram is qualitatively similar \add{for randomly distributed $m$ (\FIG{S4}), for different distributions of activity potentials (\FIG{S5}), and} for a different model in which an activated node induces a clique instead of a star~(\FIG{S\del{2}\add{6}}), modeling a group conversation event as some temporal network models do~\cite{Tantipathananandh2007,Stehle2010PhysRevE,Zhao2011}.

\noindent\emph{Discussion:}
Our analytical method shows that the presence of network dynamics boosts the prevalence (and decreases the epidemic threshold $\beta_{\rm c}$) when the concurrency $m$ is large and suppresses the prevalence (and increases $\beta_{\rm c}$) when $m$ is small, for a range of values of the network dynamic time scale $\tau$. This result lends theoretical support to previous claims that concurrency boosts epidemic spreading~\cite{Watts1992,Morris1995,Kretzschmar1996a,Bauch2000,Eames2004,Doherty2006,Eaton2011,Perra2012,Leung2015,Gurski2016}. The result may sound unsurprising because a large $m$ value implies that there exists a large connected component at any given time. However, our finding is not trivial because a large component consumes many edges such that other parts of the network at the same time or the network at other times would be more sparsely connected as compared to the case of a small $m$. \add{We confirmed that qualitatively similar results are found when the activity potentials were constructed from two empirical social contact networks (\FIG{S7}).} Our results confirm that a monogamous sexual relationship or a small group of people chatting face to face, as opposed to polygamous relationships or large groups of conversations, hinders epidemic spreading, where we compare like with like by constraining the aggregate (static) network to be the same in all cases. For general temporal networks, immunization strategies that decrease concurrency (e.g., discouraging polygamy) may be efficient. Restricting the size of the concurrent connected component (e.g., size of a conversation group) may also be a practical strategy.

Another important contribution of the present study is the observation that infection dies out  for a sufficiently large $\tau$, regardless of the level of concurrency. As shown in Figs.~3 and S6,   the transition to the ``die out'' phase occurs at values of $\tau$ that correspond to network dynamics and epidemic dynamics having comparable time scales. This is a stochastic effect and cannot be captured by existing approaches to epidemic processes on temporal networks that neglect stochastic dying out, such as  differential equation systems for pair formulation-dissolution models~\cite{Kretzschmar1996a,Bauch2000,Eames2004,Leung2015,Gurski2016} and individual-based approximations~\cite{Valdano2015,Speidel2016,Rocha2016}.  Our analysis methods explicitly consider such stochastic effects, and are therefore expected to be useful beyond the activity-driven model (or the clique-based temporal networks analyzed in the Supplemental Material) and the SIS model.

We thank Leo Speidel for discussion. We thank the SocioPatterns collaboration (http:// www.sociopatterns.org) for providing the data set. T.O. acknowledges the support provided through JSPS Research Fellowship for Young Scientists. J.G. acknowledges the support provided through Science Foundation Ireland (Grants No. 15/SPP/E3125 and No. 11/PI/1026). N.M. acknowledges the support provided through JST, CREST, and JST, ERATO, Kawarabayashi Large Graph Project.

%

\clearpage
\begin{widetext}
\begin{center}
\textbf{\large Supplemental Material for ``Concurrency-induced transitions in epidemic dynamics on temporal networks''}
\end{center}
\end{widetext}

\setcounter{equation}{0}
\setcounter{figure}{0}
\setcounter{table}{0}
\renewcommand{\theequation}{S\arabic{equation}}
\renewcommand{\thefigure}{S\arabic{figure}}

\section{Prevalence on the aggregate network}
\begin{figure}[h]
\centering
\includegraphics[width=3.4in]{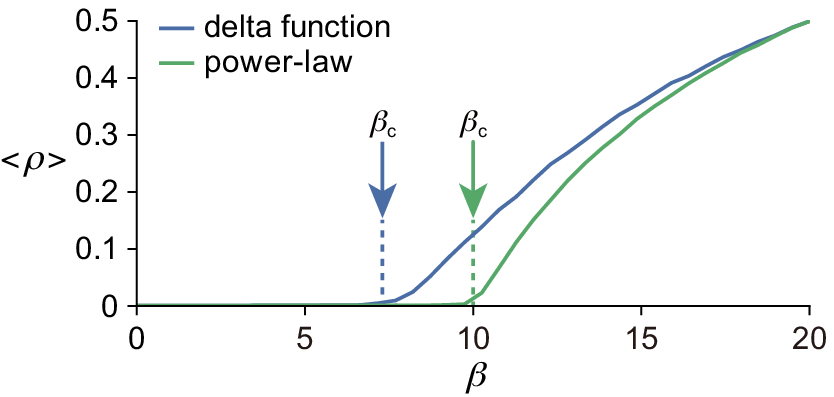}
\caption{Prevalence on the aggregate (hence static) network whose adjacency matrix is given (in the limit $N\to\infty$) by $A_{ij}^* = m(a_i+a_j)/N$~\cite{Speidel2016Supp,MasudaLambiotte2016bookSupp}. The lines represent the numerical results for the delta function (i.e., all nodes have same activity potential) and power-law activity distributions. The arrows indicate $\beta_{\rm c}=\left[  m\left(\AM+\sqrt{\AM[2]}\right)\right]^{-1}$. We set $m=5$ and $\AM=0.01$.}
\label{fs1}
\end{figure}

\begin{figure}
\centering
\includegraphics[width=3.4in]{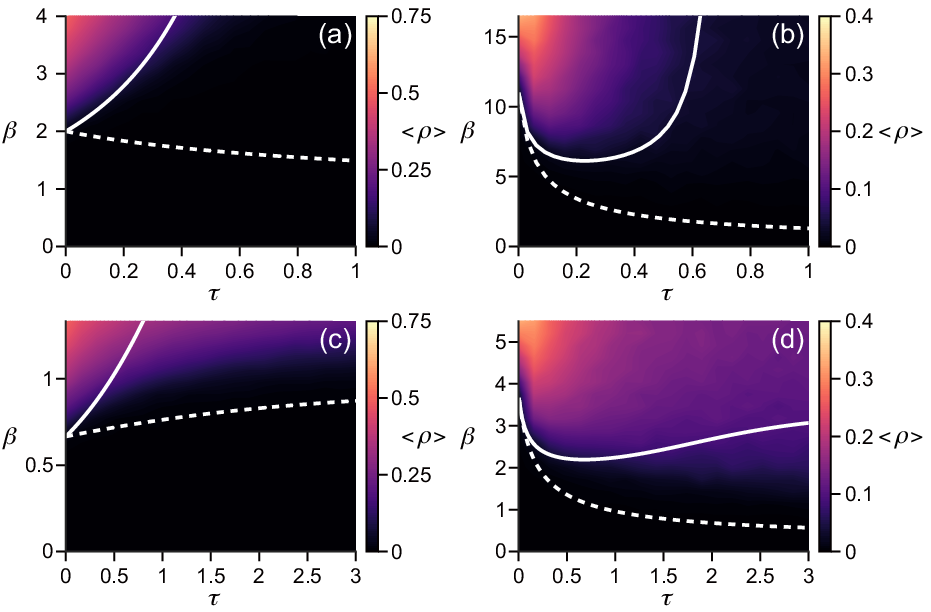}
\caption{\add{Epidemic threshold and numerically calculated prevalence when the low-activity assumption is violated. We set $m=1$ in (a) and (c), $m=10$ in (b) and (d), $p=0.5$ in (a) and (b), and $p=1.5$ in (c) and (d). The solid and dashed lines represent the epidemic threshold obtained from Eq. (8) and that obtained from the individual-based approximation, respectively. All nodes are assumed to have the same activity potential $a=0.25$ in (a), $a=0.0045$ in (b), $a=0.75$ in (c), and $a=0.0136$ in (d). We calculated the prevalence averaged over $100$ simulations after discarding the first $15000$ time steps of each simulation. We set $N=1000$ and $\Delta t=0.002$.}}
\label{fs3}
\end{figure}

\add{\section{When the low-activity assumption is violated}}
\add{Here we consider the situation in which the low-activity assumption $m^2\langle a\rangle\ll1$ is violated. When $m\ll N$, the expected number of star graphs that a star graph overlaps with is given by
\begin{equation}
  \label{eq:1}
  p=N\langle a \rangle \left[1-\left(1-\frac{m+1}{N-1}\right)^m\right] \approx m(m+1)\langle a \rangle.
\end{equation}
If $p\ll1$ is violated, a star graph would overlap with others such that the actual concurrency is larger than $m$. In the extreme case of $p\ge1$, almost all star graphs overlap with each other such that the concurrency is not sensitive to $m$. In this situation, our results overestimate the epidemic threshold because our analysis does not take into account infections across different star graphs. If $p\ge1$, the individual-based approximation describes the numerical results more accurately than our method does [Figs. \ref{fs3}(c) and \ref{fs3}(d)]. However, even at a moderately large value of $p$ $(=0.5)$, our method is more accurate than the individual-based approximation [Figs. \ref{fs3} (a) and \ref{fs3}(b)].}

\section{Derivation of $c_1$, $c_2$, $c_3$, $c_4$, and $c_5$}
We consider SIS dynamics on a star graph with $m$ leaves and derive $c_1$, $c_2$, $c_3$, $c_4$, and $c_5$. Let us denote the state of the star graph by $\{x,y,z\}~(x,y\in\{S,I\}, 0\le z\le m-1)$, where $x$ and $y$ are the states of the hub and a specific leaf node, respectively, and $z$ is the number of infected nodes in the other $m-1$ leaf nodes. Although a general network with $m+1$ nodes allows $2^{m+1}$ states, using this notation, we can describe SIS dynamics on a star graph by a continuous-time Markov process with $4m$ states~\cite{Simon2011Supp}.

We denote the transition rate matrix of the Markov process by $\mat M$. Its element $M_{\{x',y',z'\}, \{x,y,z\}}$ is equal to the rate of transition from $\{x,y,z\}$ to $\{x',y',z'\}$. The diagonal elements are given by
\begin{equation}
  M_{\{x,y,z\}, \{x,y,z\}}=-\sum_{\{x',y',z'\}\ne\{x,y,z\}}M_{\{x',y',z'\}, \{x,y,z\}}.
\end{equation}
The rates of the recovery events are given by
\begin{eqnarray} M_{\{S,y,z\},\{I,y,z\}}=&1&,\label{eq:recovery1} \\
  M_{\{x,S,z\},\{x,I,z\}}=&1&, \label{eq:recovery2} \\
  M_{\{x,y,z-1\},\{x,y,z\}}=&z&\quad(z\ge1). \label{eq:recovery3}
\end{eqnarray}
The rates of the infection events are given by
\begin{eqnarray}
  M_{\{I,S,z\},\{S,S,z\}}=&z\beta,& \label{eq:infection1}\\
  M_{\{I,I,z\},\{S,I,z\}}=&(z+1)\beta,& \\
  M_{\{I,I,z\},\{I,S,z\}}=&\beta,& \\ M_{\{I,y,z+1\},\{I,y,z\}}=&(m-1-z)\beta&\quad(z\le m-2). \label{eq:infection4}
\end{eqnarray}
The other elements of $\mat M$ are equal to $0$. Let $p_{\{x,y,z\}}(t)$ be the probability for a star graph to be in state $\{x,y,z\}$ at time $t$. Because
\begin{equation}
  \dot{\bm p}(t) = \mat M\bm p(t),
  \label{eq:p}
\end{equation}
where $\bm p(t)$ is the $4m$-dimensional column vector whose elements are $p_{\{x,y,z\}}(t)$, we obtain
\begin{equation}
  \bm p(t) = \exp(\mat Mt)\bm p(0).
  \label{eq:pt}
\end{equation}
Note that $c_1$ and $c_2$ are the probabilities with which $x=I$ at time $\tau$, when the initial state is $\{I,S,0\}$ and $\{S,I,0\}$, respectively, and that $c_3$, $c_4$, and $c_5$ are the probabilities that $y=I$ at time $\tau$, when the initial state is $\{S,I,0\}$, $\{I,S,0\}$, and $\{S,S,1\}$, respectively. Therefore, we obtain
\begin{align}
  \label{eq:cis}
  \begin{pmatrix}
    c_1 \\ c_2 \\ c_3 \\ c_4\\ c_5
  \end{pmatrix}
  =
  \begin{pmatrix}
    \sum_{y,z}\left[ \exp(\mat M\tau) \right]_{\{I,y,z\},\{I,S,0\}} \\ \sum_{y,z}\left[ \exp(\mat M\tau) \right]_{\{I,y,z\},\{S,I,0\}} \\ \sum_{x,z}\left[ \exp(\mat M\tau) \right]_{\{x,I,z\},\{S,I,0\}} \\ \sum_{x,z}\left[ \exp(\mat M\tau) \right]_{\{x,I,z\},\{I,S,0\}} \\ \sum_{x,z}\left[ \exp(\mat M\tau) \right]_{\{x,I,z\},\{S,S,1\}}
  \end{pmatrix}.
\end{align}

When $m=1$, \EQ{eq:cis} yields
\begin{eqnarray}
  &&c_1=c_3= \nonumber \\
  &&\frac{e^{-\tau}}{2} \left[e^{-\beta\tau} + e^{-\frac{1+\beta}{2}\tau}\left(\cosh\frac{\kappa\tau}{2}+\frac{1+3\beta}{\kappa}\sinh\frac{\kappa\tau}{2}\right)\right], \nonumber \\ \\
  &&c_2=c_4= \nonumber \\
  &&\frac{e^{-\tau}}{2} \left[-e^{-\beta\tau} + e^{-\frac{1+\beta}{2}\tau}\left(\cosh\frac{\kappa\tau}{2}+\frac{1+3\beta}{\kappa}\sinh\frac{\kappa\tau}{2}\right)\right], \nonumber \\
\end{eqnarray}
where $\kappa=\sqrt{\beta^2+6\beta+1}$, and $c_5$ is not defined.

When $m\gg1$, we can apply an individual-based approximation~\cite{Speidel2016Supp, Valdano2015Supp, Pastor-Satorras2015Supp}. We assume that the state of each node is statistically independent of each other, i.e.,
\begin{equation}
  p_{\{x,y,z\}}\approx P(x)P(y)P(z),
  \label{eq:independent assumption}
\end{equation}
where $P(x)$, for example, is the probability that the hub takes state $x$. We have suppressed $t$ in \EQ{eq:independent assumption}. Under the individual-based approximation, $x$ and $y$ obey Bernoulli distributions with parameters $p^{\rm MF}_1$ and $p^{\rm MF}_2$, respectively, and $z$ obeys a binomial distribution with parameters $m-1$ and $p^{\rm MF}_3$, where $\bm p^{\rm MF}\equiv (p_1^{\rm MF}, p_2^{\rm MF}, p_3^{\rm MF})^\top$ is given by
\begin{eqnarray}
  \bm p^{\rm MF}
  =
  \begin{pmatrix}
    P(x=I) \\
    P(y=I) \\
    \frac{\left<z\right>}{m-1}
 \end{pmatrix}
  =\begin{pmatrix}
 \sum_{y,z}p_{\{I,y,z\}}\\
 \sum_{x,z}p_{\{x,I,z\}}\\
 \frac{1}{m-1}\sum_{x,y,z}zp_{\{x,y,z\}}
\end{pmatrix}. \nonumber \\
  \label{pMF}
\end{eqnarray}
By substituting \EQ{eq:p} in the time derivative of \EQ{pMF}, we obtain
\begin{eqnarray}
  \dot{\bm p}^{\rm MF}
  &=&
  \begin{pmatrix}
    -p^{\rm MF}_1+\beta p^{\rm MF}_2+(m-1)\add{\beta}p^{\rm MF}_3 \\
    \beta p^{\rm MF}_1-p^{\rm MF}_2 \\
    \beta p^{\rm MF}_1(1-p^{\rm MF}_3)-p^{\rm MF}_3
  \end{pmatrix}.
  \label{eq:pdotMF}
\end{eqnarray}
If $p^{\rm MF}_{\add{3}\del{z}}\ll1$, $\bm p^{\rm MF}$ obeys linear dynamics given by
\begin{equation}
  \dot{\bm p}^{\rm MF}\approx\mat M^{\rm MF}\bm p^{\rm MF}
\end{equation}
where
\begin{equation}
  \mat M^{\rm MF}=
  \begin{pmatrix}
    -1 & \beta & (m-1)\beta \\
    \beta & -1 & 0 \\
    \beta & 0 & -1
  \end{pmatrix}.
\end{equation}
In a similar fashion to the derivation of \EQ{eq:cis}, we obtain
\begin{eqnarray}
  \begin{pmatrix}
    c_1 \\ c_2 \\ c_3 \\ c_4\\ c_5
  \end{pmatrix}
  &\approx&\begin{pmatrix}
    [\exp(\mat M^{\rm MF}\tau)]_{11} \\
    [\exp(\mat M^{\rm MF}\tau)]_{12} \\
    [\exp(\mat M^{\rm MF}\tau)]_{22} \\
    [\exp(\mat M^{\rm MF}\tau)]_{21} \\
    \frac{1}{m-1}[\exp(\mat M^{\rm MF}\tau)]_{23}
  \end{pmatrix} \nonumber \\
  &=& e^{-\tau}
  \begin{pmatrix}
    \COSH \\ \frac{1}{\SM} \SINH \\ 1+\frac{\COSH-1}{m} \\ \frac{1}{\SM}\SINH \\ \frac{1}{m}\COSHM
  \end{pmatrix}.
  \label{eq:cis for large m}
\end{eqnarray}

We estimate the extent to which \EQ{eq:cis for large m} is valid as follows. First, we need $m\gg1$, because the initial condition $p^{\rm MF}_{\add{3}\del{z}}=1/(m-1)$ should satisfy $p^{\rm MF}_{\add{3}\del{z}}\ll1$.
Second, $p^{\rm MF}_{\add{3}\del{z}}$ must satisfy
\begin{equation}
p^{\rm MF}_{\add{3}\del{z}}(\tau)\le\beta(1-e^{-\tau})+p_{\add{3}\del{z}}^{\rm MF}(0)e^{-\tau}
\end{equation}
because $p^{\rm MF}_{\add{1}\del{x}}\le1$ in \EQ{eq:pdotMF}. To satisfy $p^{\rm MF}_{\add{3}\del{z}}\ll1$, we need $\tau<1/\beta$. \add{This condition remains unchanged by re-scaling $(\tau, \beta)$ to $(c\tau, \beta/c)$.} These two conditions are sufficient for this approximation to be valid.
\add{If $m\gg1$ is violated, the individual-based approximation significantly underestimates the epidemic threshold for any finite $\tau$ because it ignores the effect of stochastic dying-out. If $\tau<1/\beta$ is violated, the approximation [dashed lines in Fig. 2(b) and (d)] underestimates the epidemic threshold because dynamics on the star graph deviate from the linear regime. In particular, the epidemic threshold obtained from the approximation [\EQ{eq:et for large m}] remains finite even in the limit $\tau\to\infty$, whereas analytical [Eq. (8)] and numerical (Fig. 2) results diverge at a finite $\tau$.}

\clearpage
\section{Derivation of Eq. (8)}
At the epidemic threshold, the largest eigenvalue of $\mat T$ is equal to unity. Let $\bm v = (v_1, v_2, \ldots)^\top$ be the corresponding eigenvector of $\mat T$. We normalize $\bm v$ such that $\sum_{j=1}^{\infty} v_j = 1$. By substituting Eq. (7) in $\bm v=\mat T\bm v$, we obtain the system of equations
\begin{eqnarray}
  v_1 &=& c_3'v_1 + c_4'\sum_{n=1}^\infty \AM[n]v_n + c_5'\sum_{n=1}^\infty \AM[n-1]v_n, \label{v1}\\
  v_2 &=& c_1'v_1 + c_3'v_2 + c_2\sum_{n=1}^\infty \AM[n-1]v_n \label{v2},\\
  v_j &=& c_1'v_{j-1} + c_3'v_j \quad (j\ge3). \label{v3}
\end{eqnarray}
Equation (\ref{v3}) gives
\begin{equation}
  v_j=\frac{q}{\AM}v_{j-1}\quad(j\ge3),
  \label{eq:vj}
\end{equation}
where
\begin{equation}
  q\equiv\frac{\AM c_1'}{1-c_3'}.
  \label{eq:q}
\end{equation}
By combining Eqs. (\ref{v2}) and (\ref{eq:vj}), we obtain
\begin{eqnarray}
  (q+r)v_1 = \AM\left[ 1- (1+qS)r \right]v_2,
\end{eqnarray}
where
\begin{eqnarray}
  r&\equiv&\frac{\AM c_2'}{1-c_3'}, \\
  S(q)&\equiv&\sum_{n=0}^\infty \frac{\AM[n+2]}{\AM^{n+2}}q^n=\frac{1}{\AM^2}\left\langle \frac{a^2}{1-\frac{a}{\AM}q} \right\rangle.
\end{eqnarray}
Because $\bm v$ is normalized, we obtain
\begin{equation}
  \bm v =
  \renewcommand{\arraystretch}{2}
  \begin{pmatrix}
    \frac{\left[ \AM-q \right]\left[ 1-(1+qS)r \right]}{r+\AM+(1+qS)\left[ q-\AM \right]r} \\
    \frac{\left[ 1-\frac{q}{\AM} \right](q+r)}{r+\AM+(1+qS)\left[ q-\AM \right]r} \\
    \frac{\frac{q}{\AM}\left[ 1-\frac{q}{\AM} \right](q+r)}{r+\AM+(1+qS)\left[ q-\AM \right]r} \\
    \frac{\left(\frac{q}{\AM}\right)^2\left[ 1-\frac{q}{\AM} \right](q+r)}{r+\AM+(1+qS)\left[ q-\AM \right]r}\\
    \vdots
  \end{pmatrix}.
  \renewcommand{\arraystretch}{1}
  \label{eq:v}
\end{equation}
Equation~(\ref{v1}) leads to
\begin{eqnarray}
    [1-s-u]v_1 = \AM\left[ sS + (1+qS)u \right]v_2,
  \label{eq:v12}
\end{eqnarray}
where, \begin{eqnarray}
  s\equiv\frac{\AM c_4'}{1-c_3'},\\
  u\equiv\frac{c_5'}{1-c_3'}.
\end{eqnarray}
By substituting \EQ{eq:v} in \EQ{eq:v12}, we obtain
\begin{eqnarray}
  f(\tau, \beta_{\rm c})&\equiv& \frac{(1-r)(1-s)-(1+q)u}{S(q)} \nonumber\\
  &-&qr-qs+qrs-q^2u-rs=0, \label{eq:f derivation}
\end{eqnarray}
which is Eq. (8) in the main text.
If all nodes have the same activity potential $a$, \EQ{eq:f derivation} is reduced to
 \begin{equation}
  f(\tau, \beta_{\rm c})=1-q-r-s-u=0.
  \label{eq:f for same activity}
\end{equation}

\add{\section{Convergence of the Maclaurin series}}
\add{ We derive the condition under which the Maclaurin series in Eq. (5) converges for any $t$ when $\beta\le\beta_{\rm c}$. First, at $t=0$, the series converges because $\bm w(0)=(p_0,0,0,...)^\top$.

  Second, consider a finite $t$. It should be noted that the series is only defined at $t$ that is a multiple of $\tau$. Because $T_{ij}=0$ $(i\ge j+2)$ in Eq. (7), we obtain
\begin{equation}
  \label{eq:3} w_n(t)=0\quad\text{ for }n\ge1+\frac{t}{\tau}.
\end{equation}
Therefore, the series converges.

Third, we consider the limit $t\to\infty$. If $\beta<\beta_{\rm c}$, because
\begin{equation}
  \lim_{t\to\infty} \langle \rho \rangle= 0,
  \label{eq:17}
\end{equation}
we obtain
\begin{equation}
  \lim_{t\to\infty}w_n(t)=0\quad\text{ for }n\ge1.
  \label{eq:16}
\end{equation}
Therefore, the series converges. For $\beta=\beta_{\rm c}$, we consider the convergence of the series when
\begin{equation}
  \lim_{t\to\infty}\bm w(t)=b\bm v,
  \label{eq:18}
\end{equation}
where $\bm v$ is the eigenvector of $\bm T$ given by \EQ{eq:v}, and $b$ is a constant. Because \EQ{eq:v} yields
\begin{equation}
  \label{eq:9}
  \lim_{j\to\infty}\frac{v_{j+1}}{v_j} = \frac{q}{\AM},
\end{equation}
the radius of convergence is equal to $\AM/q$.
To ensure convergence, we require that
\begin{equation}
  \label{eq:5}
  \max_i(a_i) < \frac{\AM}{q}.
\end{equation}
Because $c_i$ $(1\le i\le5)$ are probabilities, we obtain
\begin{align}
  c_1&\le1, \label{eq:c1le1}\\
  c_3&\le1. \label{eq:c3le1}
\end{align}
By substituting Eqs. (\ref{eq:c1le1}) and (\ref{eq:c3le1}) in the definitions of $c_1'$ and $c_2'$, we obtain
\begin{align}
  c_1'&\le 1-e^{-\tau}, \label{eq:maxc1d}\\
  c_3'&\le e^{-\tau}+m\AM(1-e^{-\tau}). \label{eq:maxc3d}
\end{align}
By substituting Eqs. (\ref{eq:maxc1d}) and (\ref{eq:maxc3d}) in \EQ{eq:q}, we obtain
\begin{equation}
  \label{eq:maxq}
  q\le \frac{\AM}{1-m\AM}.
\end{equation}
Inequalities (\ref{eq:c1le1})--(\ref{eq:maxq}) hold with equality in the limit $\beta\to\infty$.
Hence, a sufficient condition for convergence is given by
\begin{equation}
  \label{eq:condition_maxai}
  \max_i(a_i)<1-m\AM.
\end{equation}
Equation (\ref{eq:condition_maxai}) holds true in practical situations because the assumption $m^2\AM\ll1$ guarantees that $m\AM\ll1$ and $a_i$ is a probability.
}

\section{Epidemic threshold under  the individual-based approximation}
When $m\gg1$, the epidemic threshold can be obtained by the individual-based approximation~\cite{Speidel2016Supp, Valdano2015Supp, Pastor-Satorras2015Supp}. We assume that all nodes have the same activity potential $a$. By substituting \EQ{eq:cis for large m} in \EQ{eq:f for same activity}, we obtain
\begin{equation}
  \label{eq:et for large m}
  \beta_{\rm c}\approx\frac{1}{\sqrt{m}\tau}\ln\left( 1+\frac{e^\tau-1}{2\sqrt{m}a} \right).
\end{equation}
Equation (\ref{eq:et for large m}) agrees with the value derived in \cite{Speidel2016Supp}. Note that this approximation is valid only for small $\tau$ ($\tau<1/\beta_{\rm c}$).

\section{Derivation of $\tau_*$ for general activity distributions}
In the limit $\beta\to\infty$, we obtain $c_i\to1$ ($1\le i\le5$).
For general activity distributions, $f(\tau_*,\beta_{\rm c}\to\infty)=0$ leads to
\begin{eqnarray}
  \label{eq:tau_star}
  \tau_* = -\ln\left(1-\frac{b+\sqrt{b^2+4d}}{2}\right),
\end{eqnarray}
where
\begin{eqnarray}
  b &=& m\AM^2\left[ 1-m\AM \right]^{-3}\left[2-(m+1)\AM  \right]S\left(\frac{\AM}{1-m\AM}\right)\nonumber\\
  &+&m\AM\left[ 1-m\AM \right]^{-2}\left[ m+1-(m^2+1)\AM \right],\\
  d &=& m^2\AM^2\left[ 1-m\AM \right]^{-3}\left[1-(m+1)\AM  \right]S\left(\frac{\AM}{1-m\AM}\right)\nonumber\\
  &-&m^2\AM^2\left[ 1-m\AM \right]^{-2}.
\end{eqnarray}

\section{Derivation of $m_{\rm c}$ for general activity distributions}
At $m=m_{\rm c}$, an infinitesimal increase in $\tau$ from $0$ to $\Delta\tau$ does not change the $\beta_{\rm c}$ value.
For general activity distributions, by setting $\partial f/\partial\tau=0$ for $f$ given by \EQ{eq:f derivation}, we obtain
\begin{equation}
  m_{\rm c}=\frac{1+2\frac{\sqrt{\AM[2]}}{\AM}}{1-2\sqrt{\AM[2]}-2\frac{\AM[2]}{\AM}}. \label{eq:m_c}
\end{equation}

\begin{figure}
\centering
\includegraphics[width=3.4in]{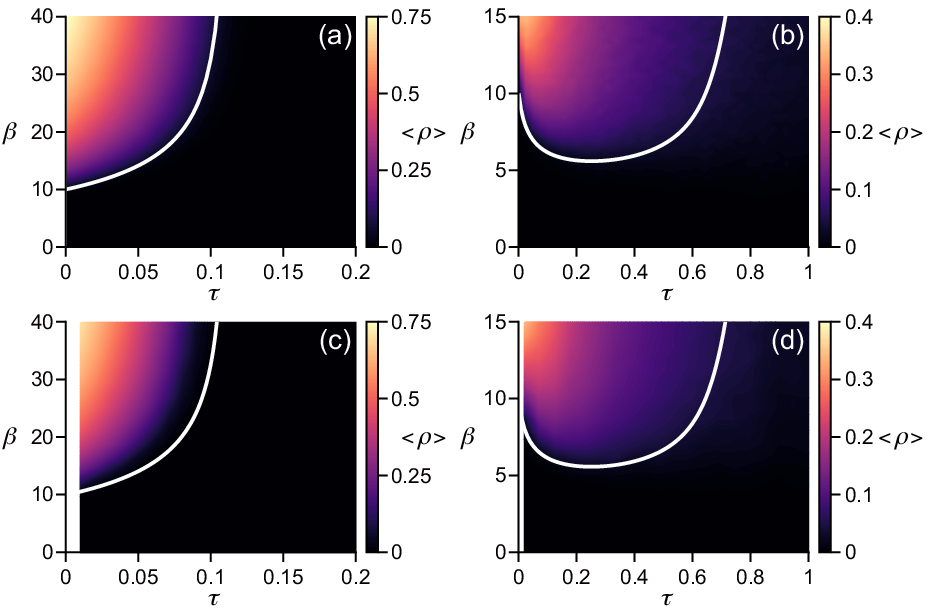}
\caption{\add{Epidemic threshold and numerically calculated prevalence for the activity-driven model with link dynamics driven by a reinforcement process~\cite{Karsai2014Supp}. We set $m=1$ in (a) and (c), and $m=10$ in (b) and (d). We used the original activity-driven model in (a) and (b) and the extended model with $c=1$ in (c) and (d). The solid lines represent the epidemic threshold obtained from Eq. (8). All nodes have $a_i=0.05$ $(1\le i \le N)$ in (a) and (c), and $a_i=0.005$ $(1\le i \le N)$ in (b) and (d). We calculated the prevalence averaged over $100$ simulations after discarding the first $15000$ time steps in each simulation. We set $N=2000$ and $\Delta t=0.002$.}}
\label{fs6}
\end{figure}

\add{\section{Activity-driven model with a reinforcement process}}
\add{We carried out numerical simulations for an extended activity-driven model in which link dynamics are driven by a reinforcement process~\cite{Karsai2014Supp}. The original activity-driven model is memoryless~\cite{Perra2012Supp}. In the extended model, an activated node $i$ connects to a node $j$ that $i$ has already contacted with probability $1/(n_i+c)$ and to a node $j$ that $i$ has not contacted with probability $c/(n_i+c)$, where $n_i$ denotes the number of nodes that node $i$ has already contacted.

  The numerically calculated prevalence is compared between the original model [Figs. \ref{fs6}(a) and \ref{fs6}(b)] and the extended model with $c=1$ [Figs. \ref{fs6}(c) and \ref{fs6}(d)]. We replicate Figs. 2(a) and 2(b) in the main text as Figs. \ref{fs6}(a) and \ref{fs6}(b) as reference. All nodes are assumed to have the same activity potential $a_i=0.05$ $(1\le i \le N)$ in (a) and (c) and $a_i=0.005$ $(1\le i \le N)$ in (b) and (d). Figure \ref{fs6} indicates that the extended model only slightly changes the epidemic threshold.
}

\add{\section{Stochastic $m$}}
\begin{figure*}
\centering
\includegraphics[width=5.5in]{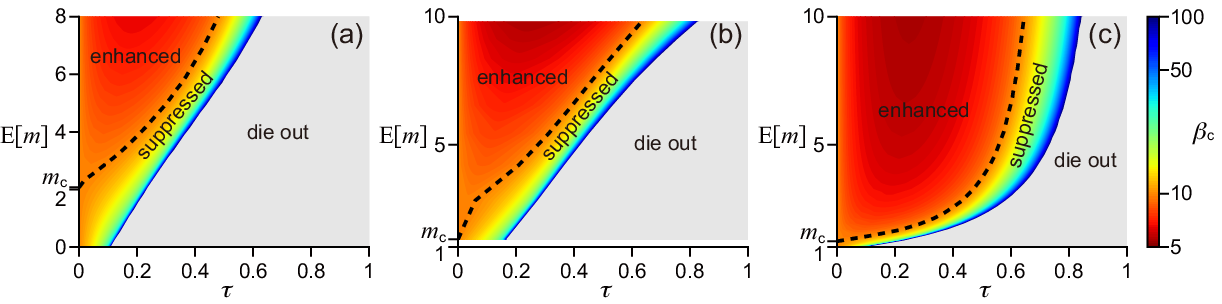}
\caption{\add{Phase diagram of the epidemic threshold when $m$ is stochastic; $m$ obeys (a) a truncated Poisson distribution $(0\le m \le m_{\max})$ and (b) a power-law distribution with exponent $3$ $(m_{\min}\le m \le m_{\max})$, and (c) a bimodal distribution in which $m$ takes $m_1$ and $m_2$ with probabilities $\tilde p$ and $1-\tilde p$, respectively. In (a) and (b), we set $m_{\max}=11$. In (a), we truncated a Poisson distribution with varying the mean between $0.01$ and $8$ to modulate $E[m]$. In (b), We vary $m_{\min}$ between $1$ to $9$. In (c), we set $(m_1, m_2)=(10, 1)$ and varied $\tilde{p}$ to modulate $E[m]$. We set $\langle k \rangle=0.1$. The dashed line represents $\tau_{\rm c}$. In the gray regions, $\beta_{\rm c}>100$.}}
\label{fs7}
\end{figure*}

\add{We consider the case in which the strength of concurrency, $m$, is not constant. To analyze this case, we change the definitions of $c_1'$, $c_2'$, $c_3'$, $c_4'$, and $c_5'$ to
  \begin{align}
    \label{eq:7}
    c_1''&={\mathrm E}[c_1-e^{-\tau}],\\
    c_2''&={\mathrm E}[mc_2],\\
    c_3''&={\mathrm E}[e^{-\tau}+m\langle a \rangle (c_3-e^{-\tau})],\\
    c_4''&={\mathrm E}[mc_4],\\
    \label{eq:8}
    c_5''&={\mathrm E}[m(m-1)\langle a\rangle c_5],
  \end{align}
  where ${\mathrm E}[\cdot]$ is the expectation with respect to the distribution of $m$. The mean degree is given by $\langle k \rangle=2a{\mathrm E}[m]$. Using Eqs. (\ref{eq:7})--(\ref{eq:8}) instead of $c_i'$ $(1\le i\le5)$, we derived the epidemic threshold in the same manner as the derivation of Eq. (8). The phase diagrams of the epidemic threshold when $m$ obeys a truncated Poisson distribution and a power-law  distribution are shown in Figs. \ref{fs7}(a) and \ref{fs7}(b), respectively. We obtain $\beta_{\rm c}=1/\langle k \rangle$ at $\tau=0$. We set the activity potential of all nodes $a=\langle k \rangle/(2{\mathrm E}[m])$ such that the epidemic threshold is the same for all ${\mathrm E}[m]$ at $\tau=0$. We numerically calculated $m_{\rm c}$ at which $\tau_{\rm c}=0$. For the power-law distribution of $m$, we cannot make ${\mathrm E}[m]$ smaller than $m_{\rm c}$ because the distribution does not have a probability mass at $m=0$ by definition. However, the phase diagrams in the case of both the truncated Poisson and power-law distributions of $m$ are qualitatively similar to the case of constant $m$.

To gain analytical insights, we calculated the phase diagrams when $m$ is equal to $m_1$ and $m_2$ with probabilities $\tilde p$ and $1-\tilde p$, respectively. We varied $\tilde p$ between $0$ and $1$. Here again, we set the activity potential of all nodes $a=\langle k \rangle/(2{\mathrm E}[m])$ such that the epidemic threshold is the same for all ${\mathrm E}[m]$ at $\tau=0$. The phase diagram [\FIG\ref{fs7}(c)] is again qualitatively similar to that found in the case of constant $m$.}

\begin{figure}
\centering
\includegraphics[width=3.4in]{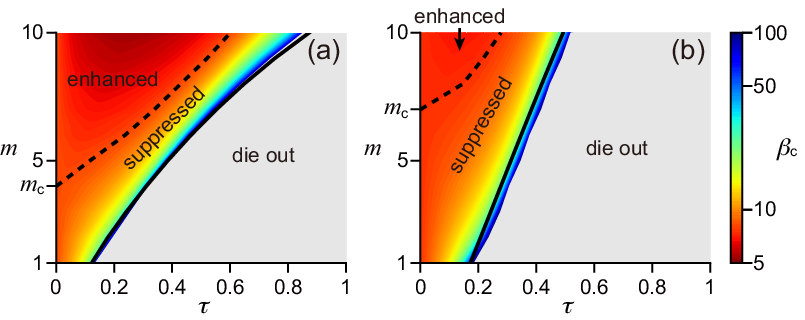}
\caption{\add{Phase diagram of the epidemic threshold when the activity potential obeys (a) an exponential distribution with a rate parameter $\lambda$ ($0\le a_i\le0.9$) and (b) a power-law distribution with exponent $2.5$ ($\epsilon\le a_i\le0.9$). We set $\langle k \rangle=0.1$ at $m=1$ and adjust the value of $\lambda$ and $\epsilon$ such that $\beta_{\rm c}$ takes the same value for all $m$ at $\tau=0$. The solid and dashed lines represent $\tau_*$ and $\tau_{\rm c}$, respectively. In the gray regions, $\beta_{\rm c}>100$.}}
\label{fs4}
\end{figure}

\add{\section{Heterogeneous activity distributions}}
\add{We analyzed the phase diagram for different distributions of activity potentials to confirm the robustness of the results shown in the main text.
We consider an exponential distribution and a power-law distribution with exponent $2.5$. We numerically calculate the epidemic threshold by solving Eq. (8) and derive $\tau_*$ and $m_{\rm c}$ from Eqs. (\ref{eq:tau_star}) and (\ref{eq:m_c}), respectively. The phase diagrams for the exponential and power-law distributions are shown in Figs. \ref{fs4}(a) and \ref{fs4}(b), respectively. These results are qualitatively similar to those found when all nodes have the same activity potential value.}

\section{Temporal networks composed of cliques}
\begin{figure}
\centering
\includegraphics[width=3.4in]{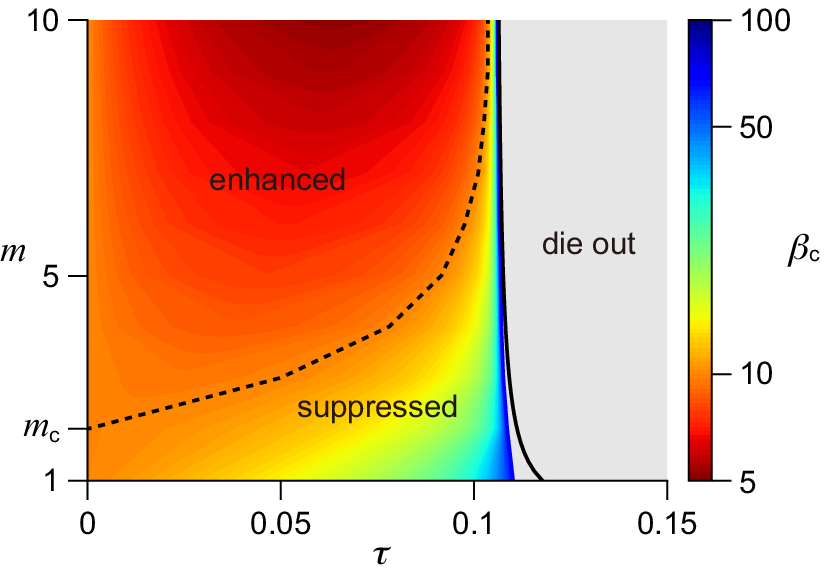}
\caption{Phase diagram of the epidemic threshold for temporal networks composed of cliques. The solid and dashed lines represent $\tau_*$~[\EQ{eq:tau star for clique}] and $\tau_{\rm c}$, respectively. All nodes are assumed to have the same activity potential given by \EQ{eq:a for clique}. We set $\langle k \rangle=0.1$.}
\label{fs2}
\end{figure}
We consider the case in which an activated node creates a clique (a fully-connected subgraph) with $m$  randomly chosen nodes instead of a star graph. This situation models a group conversation among $m+1$ people. We only consider the case in which all nodes have the same activity potential $a$. The mean degree for a network in a single time window is given by $\langle k \rangle=m(m+1)a$. The aggregate network is the complete graph. We impose $m^2a\ll1$ so that cliques in the same time window do not overlap.

As in the case of the activity-driven model, we denote the state of a clique by $\{x, y, z\}~(x,y\in\{S,I\}, 0\le z\le m-1)$, where $x$ and $y$ are the states of the activated node and another specific node, respectively, and $z$ is the number of infected nodes in the other $m-1$ nodes. The transition rate matrix of the SIS dynamics on this temporal network model is given as follows. The rates of the recovery events are given by Eqs. (\ref{eq:recovery1}), (\ref{eq:recovery2}), and (\ref{eq:recovery3}).
The rates of the infection events are given by
\begin{eqnarray}
  M_{\{I,S,z\},\{S,S,z\}}=&z\beta,& \\
  M_{\{S,I,z\},\{S,S,z\}}=&z\beta,& \\
  M_{\{I,I,z\},\{S,I,z\}}=&(z+1)\beta,& \\
  M_{\{I,I,z\},\{I,S,z\}}=&(z+1)\beta,& \\
  M_{\{S,S,z+1\},\{S,S,z\}}=&z(m-1-z)\beta&\quad(z\le m-2), \nonumber \\ \\
  M_{\{I,S,z+1\},\{I,S,z\}}=&(z+1)(m-1-z)\beta&\quad(z\le m-2), \nonumber \\ \\
  M_{\{S,I,z+1\},\{S,I,z\}}=&(z+1)(m-1-z)\beta&\quad(z\le m-2), \nonumber \\ \\
  M_{\{I,I,z+1\},\{I,I,z\}}=&(z+2)(m-1-z)\beta&\quad(z\le m-2). \nonumber \\
\end{eqnarray}
We obtain $c_i$ $(1\le i\le5)$ from $\mat M$ in the same fashion as in the case of the activity-driven model. Because of the symmetry inherent in a clique, we obtain $c_1=c_3$ and $c_2=c_4=c_5$. Therefore, \EQ{eq:f for same activity} is reduced to
\begin{equation}
f(\tau,\beta_{\rm c})=1-q-(m+1)r=0.
\end{equation}
Calculations similar to the case of the activity-driven model lead to
\begin{eqnarray} \tau_*&=&\ln\frac{1-(1+m)a}{1-(1+m)^2a}\approx\langle k \rangle, \label{eq:tau star for clique}\\
  m_{\rm c}&=&2.
\end{eqnarray}

The phase diagram shown in \FIG{\ref{fs2}} is qualitatively the same as those for the activity-driven model (\FIG{3}). Note that, in \FIG{\ref{fs2}}, we selected the activity potential value $a$ to force $\beta_{\rm c}$ to be independent of $m$ at $\tau = 0$, i.e.,
\begin{equation}
  a = \frac{\langle k \rangle}{m(m+1)}.
  \label{eq:a for clique}
\end{equation}
Although \EQ{eq:tau star for clique} coincides with the expression of $\tau_*$ for the activity-driven model [Eq. (10)], $\tau_*$ as a function of $m$ is different between the activity-driven model [a solid line in \FIG{3}(a)] and the present clique network model (a solid line in \FIG{\ref{fs2}}). This is because the values of $a$ are different between the two cases when $m\ge2$.

\add{\section{Empirical activity distributions}}
\begin{figure}
\centering
\includegraphics[width=3.4in]{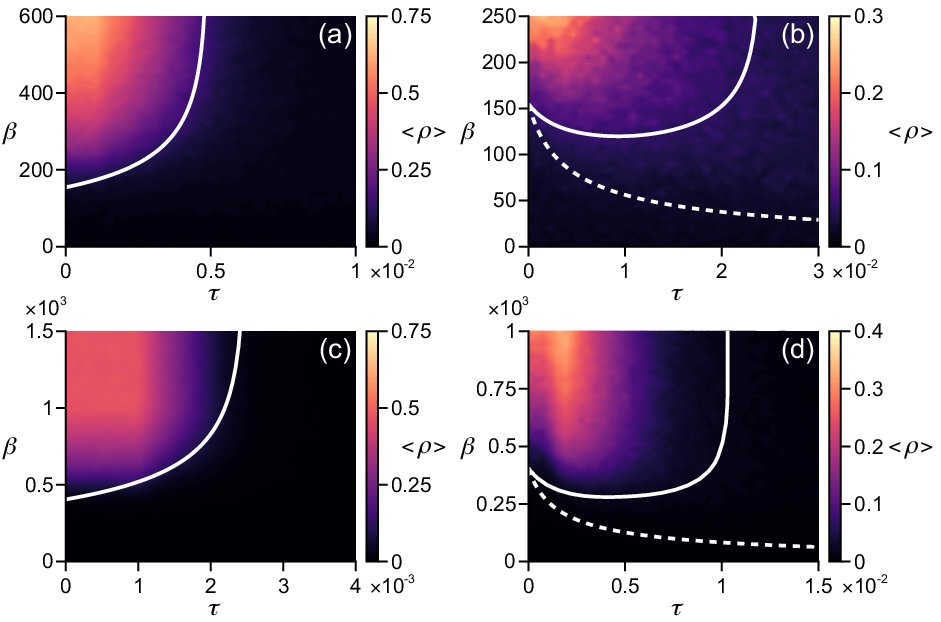}
\caption{\add{Results for activity potentials derived from empirical data. The epidemic threshold and numerically simulated prevalence are shown for $m=1$ (a),(c) and $m=10$ (b),(d). In (a) and (b), the activity potential is constructed from contact data obtained from the SocioPatterns project~\cite{Genois2015Supp}. This data set contains contacts between pairs of $N=92$ individuals measured every $20$ seconds. In (c) and (d), the activity potential is constructed from email communication data at a research institution, obtained from the Stanford Network Analysis Platform~\cite{Paranjape2017Supp}. Although the original edges are directed, we treat them as undirected. We assume that each email exchange event corresponds to a one-minute contact. We calculate the degree of each node per minute averaged over time, denoted by $\langle k_i \rangle$, and define the activity potential as $a_i=\left[\langle k_i \rangle-\langle k \rangle/2 \right]/m$. In (c) and (d), we used $N=439$ individuals satisfying $a_i>0$ (some individuals exchanged few emails such that $a_i<0$). We set $\Delta t=0.001$.}}
\label{fs5}
\end{figure}
\add{The epidemic threshold and prevalence when $F(a)$ is constructed from empirical contact data at a workplace, obtained from the SocioPatterns project \cite{Genois2015Supp}, are shown in Figs. \ref{fs5}(a) and \ref{fs5}(b) for $m=1$ and $m=10$, respectively. The results for $F(a)$ constructed from email communication data at a research institution, obtained from the Stanford Network Analysis Platform \cite{Paranjape2017Supp}, are shown in Figs. \ref{fs5}(c) and \ref{fs5}(d) for $m=1$ and $m=10$, respectively. These results are qualitatively similar to those shown in \FIG 2.}

%

\end{document}